\documentstyle[epsfig]{aipproc}
\input epsf

\newcommand{\qtwom}{\mbox{$Q^2 \;$}}

\newcommand{\gpm} {${\rm \gamma-p}$}
\newcommand{\xl} {${\rm x_L}$}

\newcommand{\bi} {\begin{itemize}}
\newcommand{\ei} {\end{itemize}}

\newcommand{\bc} {\begin{center}}
\newcommand{\ec} {\end{center}}

\begin{document}
\title{Leading Baryons at Low \xl $\;$ in DIS and Photoproduction at ZEUS}

\author{Nicol\`{o} Cartiglia}
\address{Columbia University, Nevis Laboratories, \\
136 South Broadway, Irvington N.Y., 10533 USA}

\maketitle

\begin{abstract}
Results obtained by the  ZEUS collaboration on leading baryon production in the proton fragmentation region are presented. The reaction ${\rm \gamma p \rightarrow N X}$,  with N a proton  or a neutron, is examined  both at low and high photon virtuality. 
\end{abstract}

\section*{Definitions and Kinematics}
\begin{figure}[htb] 
\centerline{\epsfig{file=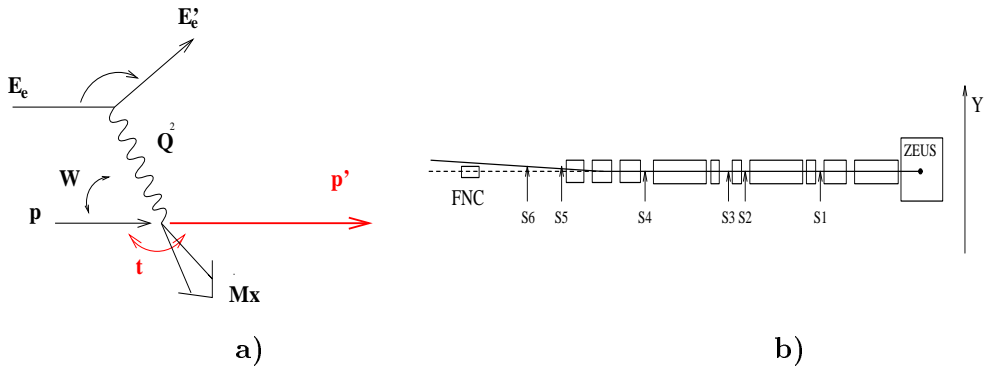,height=2.in,width=5.5in}}
\vspace{20pt}
\caption{ a$)$ Diagram for the reaction ${\rm ep \rightarrow N X}$ b$)$ Schematic of the ZEUS far forward detector system. S1,..,6 are the six LPS stations}
\label{fig:kin_for}
\end{figure}

At the HERA collider at DESY (Hamburg, Germany), 820 GeV protons collide
with 27.5 GeV electrons or positrons. 
In  Fig.~\ref{fig:kin_for}(a), a diagram of the reaction ${\rm e p \rightarrow  N X}$
 is shown. A photon ${\rm \gamma^*}$ with virtuality ${\rm q^2 =-}$\qtwom is emitted at the electron vertex . 
${\rm W^2=(q+p)^2}$ is used to denote  the centre of mass energy of the virtual photon-proton (\gpm) system and $x, y$  are the standard Bjorken variables. The mass of the proton is denoted by ${\rm m_p}$.

In addition, leading baryons (LB)\footnote{ LP, LN stand for leading proton and neutron respectively}
 are described by two other variables: t, the square of the
four-momentum transfer at the proton  vertex, defined as ${\rm
t=(p-p^{\prime})^2\simeq -\frac{ (p^\prime_\perp)^2}{x_L} - m_p^2 \frac{(1-x_L)\
^2}{x_L} }$ and  \xl, the fraction of the incoming proton momentum carried by the leading baryon, defined as ${\rm  x_L = p^{\prime}_z/p_z}$.

In the following, deep inelastic scattering (DIS)  events are defined as those having  photon virtuality  ${\rm Q^2 > 4 }$ GeV$^2$ and the energy of the  scattered electron ${\rm E_e > 8}$ GeV,  while photoproduction (PHP) events have ${\rm Q^2 < 0.02 }$ GeV$^2$ and ${\rm 12 < E_e < 18}$ GeV.

The ZEUS detector\cite{ref:zeus}, one of the two general purpose detectors  at HERA, is instrumented with high resolution
calorimeters and tracking chambers. In the far forward direction\footnote{ Following the  HERA convention,  angles are  measured with respect  to the proton beam direction. Therefore `forward direction' is used to indicate the direction of the proton beam. Angles are often expressed using the pseudorapidity $\eta$  defined as:  ${\rm \eta=-ln(tan( \theta/2))}$. The central ZEUS calorimeter measures up to ${\rm \eta = 4.2}$. }, it is equipped with a  lead scintillator 
 calorimeter\cite{ref:FNC} (forward neutron calorimeter, FNC),
  located 106 m downstream of the nominal interaction point, and with a leading proton spectrometer\cite{ref:LPS} (LPS), consisting of silicon strip detectors operating inside movable roman pots located in six stations between 23 and 90 m downstream  of the interaction region, Fig~\protect\ref{fig:kin_for} b). 

\section*{Properties of  events with a leading baryon}

\begin{figure}[htb] 
\centerline{\epsfig{file=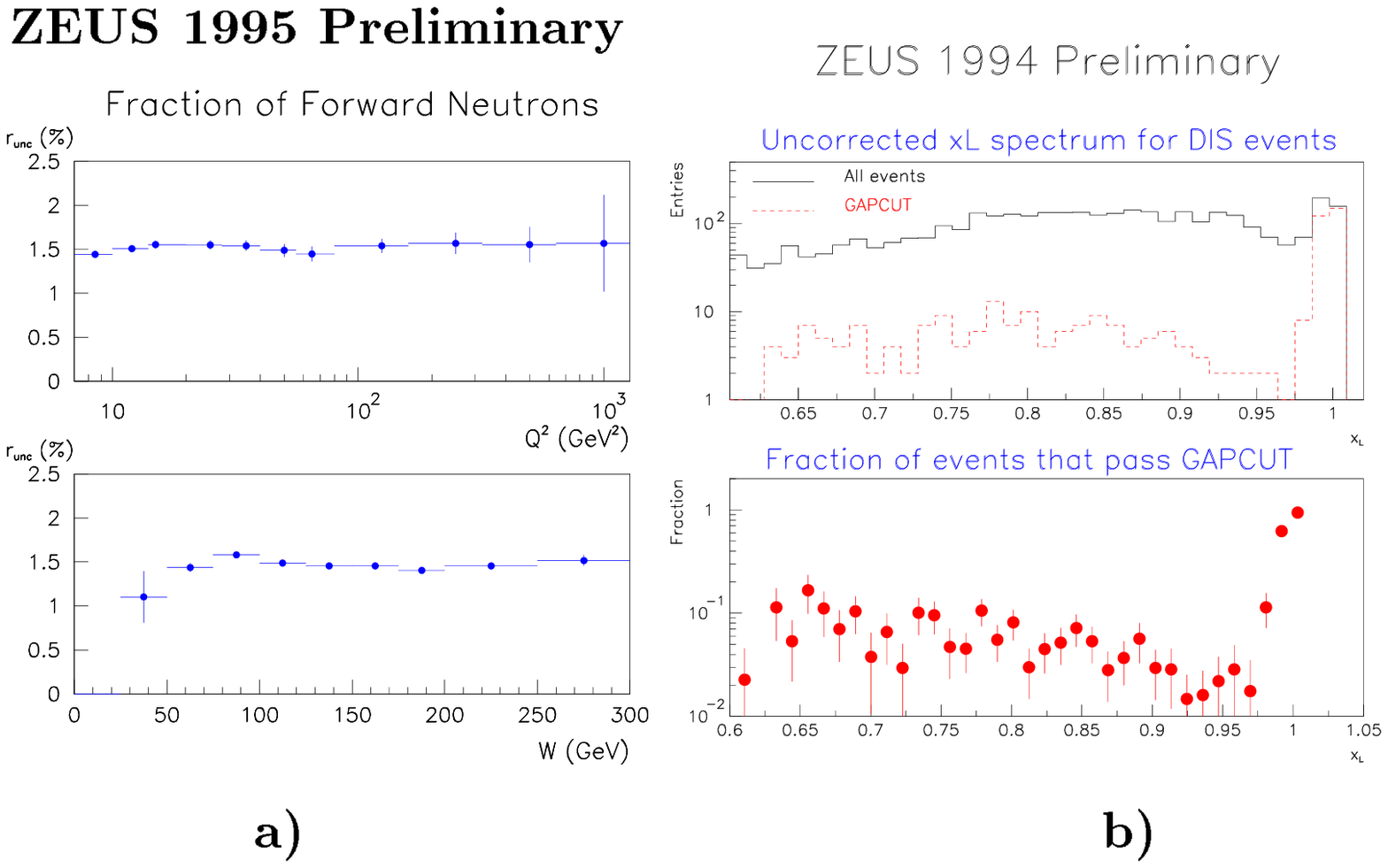,height=3in,width=5.5in}}
\vspace{20pt}
\caption{ a$)$ Fraction of DIS events with a LN in the final state vs W and \qtwom $\;$ b) \xl $\;$ spectrum for all DIS events, for GAPCUT events and the ratio of the two spectra.}
\label{fig:dis_prop}
\end{figure}

 Fig~\protect\ref{fig:dis_prop} a) shows the  fraction of DIS events with a LN  with ${\rm E_n > 200 GeV}$ as a function of \qtwom and W (the ratio is uncorrected for detector acceptance). As can be seen from these plots, and  analogous ones for events with a LP, the fraction of DIS events with a LB is independent of x, y, and \qtwom.

To estimate the fraction  of events with a LP   generated by pomeron exchange, the `GAPCUT'  selection method was introduced. An event is accepted by GAPCUT if either the  pseudorapidity of the most forward energy deposition in the central detector is less than 1.8  or  a pseudorapidity  gap of at least 1.5 units with its forward edge between ${\rm 2.5 < \eta < 4}$ is present. According to MC studies, GAPCUT has an efficiency of 40-60\% for pomeron mediated events, while it  rejects non pomeron events with an efficiency ${\rm  > 99 \%}$.
  Fig~\protect\ref{fig:dis_prop} b) shows the  \xl $\;$ spectrum for all DIS events, for those events that pass GAPCUT and the ratio of the two spectra. It is clear from this  picture that only a small fraction of events with ${\rm 0.6 < x_L <0.9}$ is generated by pomeron exchange.

\begin{figure}[htb] 
\centerline{\epsfig{file=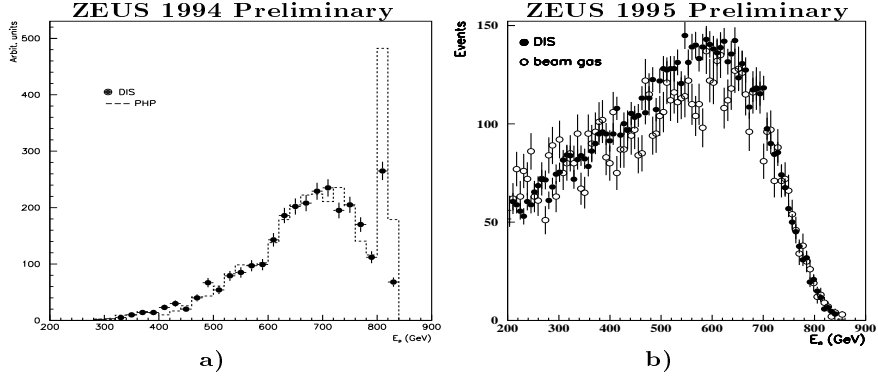,height=2.in,width=5.in}}
\vspace{20pt}
\caption{ a$)$ LP spectra measured in DIS and PHP events b) LN spectra measured in DIS and beam gas events}
\label{fig:xlfnc_lps}
\end{figure}

LB production has been studied both at high and low \qtwom.  Fig~\protect\ref{fig:xlfnc_lps} a) shows the \xl $\;$ spectra for LP measured  in DIS and PHP events while  Fig~\protect\ref{fig:xlfnc_lps} b) shows  the \xl $\;$ spectra for LN measured in DIS events and in interactions of the proton beam with gas in the beam-pipe. Both a) and b) are not corrected for LPS or FNC acceptance so they cannot be directly compared with each other. In both figures, the relative normalization of the two spectra is arbitrary. The similarity of the two spectra in each figure is  remarkable, suggesting a production mechanism  that does not depend on the type of interaction.

\begin{figure}[htb] 
\centerline{\epsfig{file=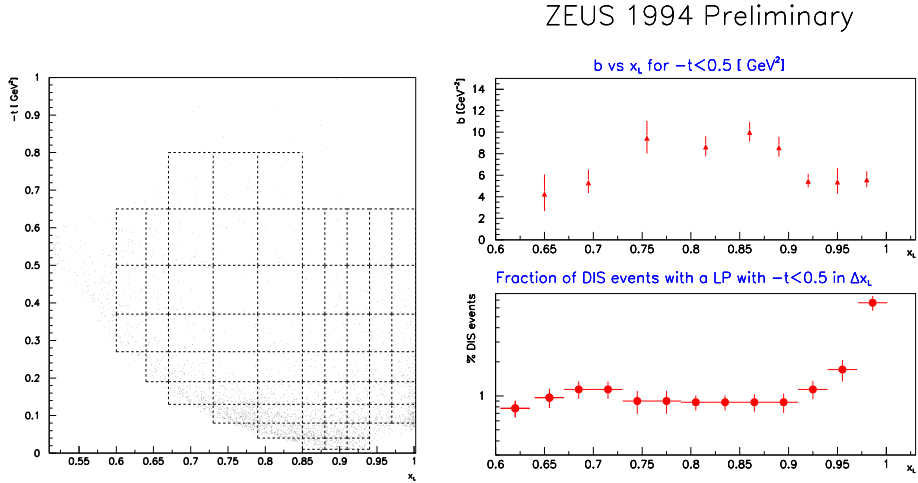,height=2.6in,width=5.5in}}
\vspace{20pt}
\caption{ Left side: \xl-t bins used in the fits. Right side: (top) measured slope parameter b in bins of \xl $\;$, (bottom) fraction of DIS events with a LP in a given ${\rm \Delta x_L}$ bin.}
\label{fig:fig4}
\end{figure}

For LP, the \xl $\;$ spectrum was divided into ten bins, and in each bin the t distribution was fitted using a simple exponential ${\rm ( dN/dt \propto e^{bt}) }$. Three different   t ranges have been tried
  for the fit (${\rm 0.08 <-t<0.5\; GeV^2, -t<0.5\ GeV^2,\; all\; available\; bins}$) and no systematic effect on the result was observed.  Fig~\protect\ref{fig:fig4}, on the left, shows the bins available for  the fit. On the right, the top figure shows the measured `b' values as a function of \xl $\;$ while on the bottom the acceptance corrected fraction of DIS events with a leading proton in a given ${\rm  x_L}$ bin is shown. This plot  shows that 8-10~\% of DIS events have a LP in the interval ${\rm  -t < 0.5 \; GeV^2  \; and \; 0.6 < x_L < 0.9 }$. For LN\cite{ref:ln}, in the interval ${\rm -t < 0.5 \; GeV^2 \; and \; x_L>0.5 }$, this fraction is  ${\rm 9.1^{+3.6}_{-5.7} }$~\%. These two measurements, even if both  have large errors, indicate that events with a  LB represent a significant fraction of all DIS events and that the amount of LP and LN is comparable. 

\begin{figure}[htb] 
\centerline{\epsfig{file=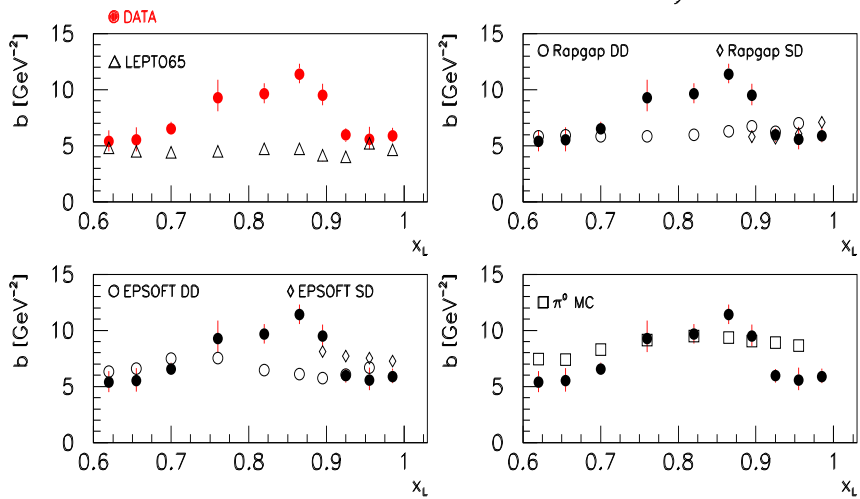,height=3in,width=5.5in}}
\vspace{20pt}
\caption{Comparison DATA-MC for b in bins of \xl}
\label{fig:fig5}
\end{figure}

The values of b in bins of \xl $\;$ are compared to different MC models in Fig~\protect\ref{fig:fig5}. 
LEPTO6.5\cite{ref:LEPTO}, which generates LB  via `soft colour interaction' and  is proposed as a model capable to explain all aspects of DIS events, including LB production and diffraction, fails to reproduce the data. LEPTO6.5 also predicts a fraction of GAPCUT events much higher than measured. RAPGAP\cite{ref:rap} and EPSOFT\cite{ref:ep}\footnote{EPSOFT models  diffractive processes as  soft hadronic collisions }, which both describe diffractive scattering, have been used to simulate both single (SD) and double diffraction (DD)(in DD  the proton also breaks up and a leading proton with \xl$<$1 can be generated in the fragmentation). These models however cannot be compared to the data over the whole \xl $\;$ range since they simulate only the events produced by pomeron exchange which are, according to  Fig~\protect\ref{fig:dis_prop} a), only  a small fraction of the total  sample at low \xl. For \xl$>$0.95, where pomeron exchange dominates, data and the various  MC predictions agree.
 RAPGAP was also used to generate `pion exchange' events, in which the virtual photon scatters off a pion  generated at the proton vertex via the processes ${\rm p\rightarrow p \pi^0 \;\; or \;\; p\rightarrow n \pi^+  }$. According to \cite{ref:kop}, pion exchange is expected to dominate the cross section for LB production in the interval ${\rm 0.7<x_L<0.9}$. Fig~\protect\ref{fig:fig5} d) shows that this model
agrees well with the data in this interval.

The value of the slope parameter `b' was also  measured  in two large \xl $\;$ intervals, ${\rm 0.6< x_L<0.75 \; and \;  0.75< x_L < 0.9}$, for the subset of events that pass GAPCUT. In both intervals, the value ${\rm b_{GAPCUT} }$ is consistent with the values of the  LP measured in all DIS events as if the fact that the interaction is diffractive did not influence the properties of the LP.

\end{document}